\documentstyle[12pt,epsf]{article}

\title{Theoretical Necessity of an External Scalar Field in the Kaluza-Klein
Theory (I)}

\author{J.P.  Mbelek and M.  Lachi\`eze-Rey \\ Service d'Astrophysique, C.E.
Saclay \\ F-91191 Gif-sur-Yvette Cedex, France}

\begin{document} \maketitle \baselineskip=8mm

\begin{abstract} We show that the principle of least action is generally
inconsistent with the usual Kaluza-Klein program, the higher dimensional
Einstein-Hilbert action being unbounded from below.  This inconsistency is also
present in other theories with higher dimensions like supergravity.  Hence, we
conclude to the necessity of an external scalar field to compensate this flaw.
\end{abstract}

\section{Introduction} Landau and Lifshitz \cite{Landau} have shown that the
principle of least action is meaningful for Einstein general relativity only if
the gravitation constant, G, is positive.  First they subject the metric tensor,
$(g_{\mu\nu})_{\mu, \nu = 0, 1, 2, 3}$, to four (the number of spacetime
coordinates, $x^{\mu}$ ; $x^{0} = ct$) conditions which suppress the liberty of
a gauge choice.  With these restrictions, the variations $\delta g_{\mu\nu}$
correspond really to a change of the gravitational field and not to a
coordinates transformation.  They make the choice \begin{equation} \label{LL
conditions} g_{0k} = 0 \,\,\,\,\,\,\,\,\,\,and
\,\,\,\,\,\,\,\,\,\,\det{(g_{kl})} = constant , \end{equation} where $k, l = 1,
2, 3$.  The only relevant terms of the Einstein-Hilbert action, $-
\,\frac{1}{2\chi} \,\sqrt{-g} \,R$, involving time derivatives of the metric
tensor are thus of the form (the signature of the metric will be + - - - and $g
= \det{(g_{\mu\nu}})$ throughout this paper) :  \begin{equation} \label{LL}
\frac{1}{8\chi} \,g^{00} \,g^{kl} \,g^{mn} \,\frac{\partial g_{km}}{\partial
x^{0}} \,\frac{\partial g_{ln}}{\partial x^{0}} \,\sqrt{- g} \end{equation}
which reduce to \begin{equation} \label{LL bis} \frac{1}{8\chi} \,g^{00}
\,(\frac{\partial g_{kl}}{\partial x^{0}})^{2} \,\sqrt{- g} \end{equation} in
local spatial cartesian coordinates ($g_{kl} = g^{kl} = {\delta}_{kl}$).  Now,
since $g^{00} = 1/g_{00} > 0$ ($g_{\mu\nu} g^{\mu\nu} = 4$ and $g_{kl} g^{kl} =
3$), the sign of the quantity (\ref{LL bis}) above is evident and well defined.
If the Einstein gravitational constant $\chi = 8\pi G/c^{4}$ were negative,
sufficiently rapid changes of the ${g_{kl}}$ with respect to time would lead to
arbitrarily large negative values of the action and thus to instabilities
without limit.  Hence, Landau and Lifshitz conclude that G should be positive.
Indeed, as emphasized by Noerdlinger \cite{Noerdlinger}, it would not be
possible to replace the principle of least action with a principle of greatest
action since physical examples, such as the free motion of a test particle,
satisfy the principle of least action\footnote{Nevertheless, let us recall that
the requirement of an extremal action is sufficient to derive the field
equations.}.\\\\In the Brans-Dicke theory \cite{Brans, Dicke}, a scalar field,
$\phi$, enters in the metric and the action may be expressed in two possible
frames conformally related one to the other:  the Jordan-Fierz frame and the
Einstein-Pauli frame\footnote{In the Jordan-Fierz frame the action terms for the
ordinary matter (other than the Brans-Dicke scalar and non-gravitational) take
the general relativistic form whereas in the Einstein-Pauli frame the
gravitational term of the action is of the Einstein-Hilbert form.}.  Noerdlinger
\cite{Noerdlinger} has shown that a similar argument applies in the Jordan-Fierz
frame to the additive term \begin{equation} \label{BD} \frac{\omega}{\phi}
\,(\frac{\partial \phi}{\partial x^{0}})^{2} \,\sqrt{- g}, \end{equation} which
similarly requires the posivity of the Brans-Dicke coupling
constant\footnote{This was first put forward by Brans and Dicke from physical
considerations concerning the positivity of the contribution to the inertial
reaction ({\it i.e.}  to the Brans-Dicke scalar) from nearby matter
\cite{Brans}.}, $\omega$, with the implicit assumption of a positive defined
scalar field.  Let us notice that the positivity of $\phi$ (see Hawking and
Ellis \cite{Hawking}) also follows from the argument of Landau and Lifshitz,
though not emphasized by Noerdlinger.  Moreover, the argument of Landau and
Lifshitz holds both in Einstein-Pauli frame and Jordan-Fierz frame,
independently of their respective physical significance.  Thus, since the
constraint on the allowed values of $\omega$ implied by the argument of Landau
and Lifshitz is stronger in Jordan-Fierz frame ($\omega > 0$) than in
Einstein-Pauli frame ($2\omega + 3 > 0$), clearly it is the former that should
be considered as relevant for our purpose.  These results seem largely
unknown\footnote{The sole quotation of Noerdlinger'$^{s}$ paper we have found
yet in the literature is from a paper of Nordtvedt \cite{Nordtvedt} that goes
back to 1970 (see also Ni \cite{Ni}).}.  Indeed, even today, one often finds in
the literature some studies on the Brans-Dicke theory with $\omega < 0$ although
the variational principle is postulated.  In particular, the low-energy
effective action of string theory, in the graviton-dilaton sector, can be given
in the form of an effective Brans-Dicke action with coupling constant $\omega =
-1$ which, according to the previous argument, would lead to devastating
instabilities.  This feature of string theory has raised no discussion in the
literature hitherto.  Let us emphasize that the argument of Landau and Lifshitz,
based on the requirement of a lower bound for the action, is stronger than the
usual argument based on the weak energy condition.  Quite often \cite{Lidsey,
Maggiore}, an argument based on the weak energy condition is invoked which leads
to the weaker constraint $2\omega + 3 > 0$.  This constraint is compatible with
the low-energy string limit $\omega = -1$.

\section{The case of the Kaluza-Klein theory} Here we apply an analogous to
Landau and Lifshitz argument to Kaluza-Klein theory.  First, let us consider the
case of a classical real scalar field, $\phi$, minimally coupled to gravity.  As
one knows, the action of the system writes \begin{equation} \label{minimal
scalar} S = \int \sqrt{- g} \,(- \,\frac{R}{{\kappa}^{2}} \,+ \,\frac{1}{2}
\,{\partial}^{\mu} \phi \,{\partial}_{\mu} \phi - U - J \,\phi) \,d^{4}x,
\end{equation} where $U = U(\phi)$ denotes the potential of the $\phi$-field, $J
= J(x^{\mu})$ is its source term and we have set $\kappa = \sqrt{2\chi}$.
Clearly, the argument of Landau and Lifshitz applies without inconsistency
because the signs of the Ricci scalar, R, and of the kinetic term of the
$\phi$-field are opposit.  Furthermore, this remains true for other kinds of
covariant couplings (in particular for a conformal coupling) and for a complex
scalar field.\\\\The action for the classical five dimensional spacetime
Kaluza-Klein theory reads (hereafter, any quantity carrying a hat is five
dimensional, the other notations are obvious) \begin{equation} \label{KK action}
S_{KK} = - \,\int \sqrt{- \hat{g}} \,\frac{\hat{R}}{{\hat{\kappa}}^{2}}
\,d^{5}x.  \end{equation} Relation (\ref{KK action}) can be expressed, after
dimensional reduction, either in the Jordan-Fierz frame or in the Einstein-Pauli
frame.  One passes from the point of view of the Jordan-Fierz frame to that of
the Einstein-Pauli frame by the conformal transformation \begin{equation}
\label{conformal transformation} {\hat{g}}_{AB} \rightarrow
{\tilde{\phi}}^{-1/3} \,{\hat{g}}_{AB} \end{equation} and the following
redefinition of the $\phi$-field \begin{equation} \label{redefinition of phi}
{\hat{g}}_{44} = {\phi}^{2} \rightarrow {\hat{g}}_{44} = \tilde{\phi},
\end{equation} where the ${\hat{g}}_{AB}$ (resp.  the ${\hat{g}}^{AB}$) denote
the covariant (resp.  contravariant) components of the five dimensional metric ;
$A, B$ (resp.  $M, N$) $= 0, 1, 2, 3, 4$.  One gets,

\begin{enumerate}

\item in the Jordan-Fierz frame \cite{Thiry}, \cite{Overduin} :
\begin{equation} \label{KK action Jordan} S_{KK} = - \,\int \sqrt{- g} \, \phi
\,(\frac{R}{{\kappa}^{2}} \,+ \,\frac{1}{4} \, {\phi}^{2} \,F^{\mu\nu}
\,F_{\mu\nu} \,+ \,\frac{2}{{\kappa}^{2}} \,\frac{{\partial}^{\mu} \phi
\,{\partial}_{\mu} \phi}{{\phi}^{2}}) \,d^{4}x, \end{equation}

\item in the Einstein-Pauli frame \cite{Overduin}, \cite{Iyer}, \cite{Collins},
\cite{Appelquist} :  \begin{equation} \label{KK action Pauli} S_{KK} = - \,\int
\sqrt{- \tilde{g}} \,(\frac{\tilde{R}}{{\kappa}^{2}} \, \,+ \,\frac{1}{4} \,
\tilde{\phi} \,F^{\mu\nu} \,F_{\mu\nu} \,+
\,\frac{1}{6{\kappa}^{2}{\tilde{\phi}}^{2}} \,{\partial}^{\mu} \tilde{\phi}
\,{\partial}_{\mu} \tilde{\phi}) \,d^{4}x, \end{equation} where ${\kappa}^{-2} =
\int {\hat{\kappa}}^{-2} \,dx^{4}$ and the ${F_{\mu\nu}} $ (resp.  ${F^{\mu\nu}}
$) are the covariant (resp.  contravariant) components of the electromagnetic
strength tensor :  $F_{\mu\nu} = {\partial}_{\mu} A_{\nu} - {\partial}_{\nu}
A_{\mu}$ (resp.  $F^{\mu\nu} = {\partial}^{\mu} A^{\nu} - {\partial}^{\nu}
A^{\mu}$).

\end{enumerate}

Relations (\ref{KK action Pauli}) and (\ref{KK action Jordan}) are equivalent
only with respect to the ground state of the $\phi$-field the vacuum expectation
value of which is $<\phi> \,= 1$.\\\\As one can see (by comparing relations
(\ref{KK action Pauli}) and (\ref{minimal scalar}) or relations (\ref{KK action
Jordan}) and (\ref{minimal scalar})), in both frames the sign of the kinetic
part of the $\phi$-field in the Kaluza-Klein lagrangian density is negative.
Hence, unless one considers the limiting case $\phi =$ contant as did indeed
Kaluza \cite{Kaluza} and Klein \cite{Klein}, applying analogously the argument
of Landau and Lifshitz to relation (\ref{KK action Pauli}) or (\ref{KK action
Jordan}) reveals an inconsistency.  Indeed, as first pointed out by Thiry
\cite{Thiry} and Jordan \cite{Jordan}, it is well known that the case $\phi =$
contant is too restrictive requiring the strict equality of the magnitudes of
the magnetic field and the electric field (up to the factor c, velocity of light
in the vacuum).  Thus, we conclude that the classical Kaluza-Klein theory is
unstable, in the sense that its action turns out to be unbounded from below.

seems to hold whatever the number of dimensions of the internal space may be
(see the expression of the action obtained by Cho and Keum \cite{Cho} for the
four-dimensional Einstein-Yang-Mills theory after dimensional reduction).  To
our knowledge, this point has never been discussed hitherto.  Perhaps this comes
from the belief that, when one neglects the electromagnetic field, the
Kaluza-Klein theory reduces to the special case $\omega = 0$ of the Brans-Dicke
theory (which is true but needs further discussion).  Also, errors on the sign
of the kinetic term of the Kaluza-Klein scalar are frequent in the literature,
different from one author to another, even using the same signature of the
metric.  In order to be confident on the latter point, we have carried out the
calculations of the Kaluza-Klein action both in Jordan-Fierz and Einstein-Pauli
frames (see appendix).  In the Jordan-Fierz frame, we have found the same
expression for the five dimensional Ricci scalar as Thiry \cite{Thiry} (see also
Lichnerowicz \cite{Lichnerowicz}) who used the Cartan method and an orthonormal
mooving frame.  Furthermore, let us notice that the inspection of the
supergravity lagrangian density (see Bergshoeff {\it et al.}  \cite{Bergshoeff})
shows that the same remark and the same conclusion as for the Kaluza-Klein
theory may be made for supergravity in ten dimensions.

\section{Conclusion} We have shown the inconsistency of the purely geometrical
Kaluza-Klein program, due to the sign of the kinetic term of the scalar field
which leads to an action unbounded from below and thus to instabilities.  Thus
we claim that any theory which leads to this form, as low energy limit, must be
rejected unless an efficient stabilizing mechanism is provided.  In a
forthcoming paper, we will propose such a solution, restoring a lower bound for
the action in the framework of the Kaluza-Klein five dimensional unification
theory.  The perturbing negative sign in the Kaluza-Klein action is compensated
by introducing an additional real (external) scalar field, which is minimally
coupled to gravity.  We do not claim that this is the only possibility but this
would bring some more clarification on the central role of the Higgs field in
particle physics or inflaton in cosmology.\\

\section{Appendix :  Calculation of the five dimensional Kaluza-Klein action} We
adopt the signature + - - - \,.  The five dimensional Ricci scalar reads
\begin{equation} \label{Ricci 5D} \hat{R} = {\hat{g}}^{AB} \,( \,{\partial}_{N}
{\hat{\Gamma}}^{N}_{AB} - {\partial}_{B} {\hat{\Gamma}}^{N}_{AN} +
{\hat{\Gamma}}^{N}_{AB} \,{\hat{\Gamma}}^{M}_{NM} - {\hat{\Gamma}}^{M}_{AN}
\,{\hat{\Gamma}}^{N}_{BM} \,).  \end{equation} This involves (see Landau and
Lifshitz \cite{Landau}) :  \begin{equation} \label{EinsteinHilbert 5D} \sqrt{ -
\hat{g}} \,\hat{R} = \sqrt{ - \hat{g}} \,\hat{G} + \hat{D}, \end{equation} with
\begin{equation} \label{G quantity 5D} \hat{G} = {\hat{g}}^{AB} \,(
\,{\hat{\Gamma}}^{M}_{AN} \,{\hat{\Gamma}}^{N}_{BM} - {\hat{\Gamma}}^{N}_{AB}
\,{\hat{\Gamma}}^{M}_{NM} \,) \end{equation} and, on account of the cylinder
condition, \begin{equation} \label{pseudodivergence} \hat{D} = {\partial}_{\nu}
(\sqrt{ - \hat{g}} \,{\hat{g}}^{AB} \,{\hat{\Gamma}}^{\nu}_{AB}) -
{\partial}_{\beta} (\sqrt{ - \hat{g}} \,{\hat{g}}^{A\beta}
\,{\hat{\Gamma}}^{N}_{AN}).  \end{equation} Analogously to the conditions
(\ref{LL conditions}) set by Landau and Lifshitz for the three-dimensional space
in spacetime, we compute the quantities (\ref{G quantity 5D}) and
(\ref{pseudodivergence}) assuming the choice of a set of spacetime coordinates
that respects the following conditions \begin{equation} \label{LL conditions in
5D} {\hat{g}}_{4\mu} = 0 \,\,\,\,\,\,\,\,\,\,and
\,\,\,\,\,\,\,\,\,\,\det{(g_{\mu\nu})} = constant.  \end{equation} This choice
strongly simplifies the calculations and thus avoids many errors.  In
particular, it is straightforward that\footnote{Relation (\ref{LL conditions in
5D}) does not involve the cancellation of the derivatives of the
${{\hat{g}}_{4\mu}}$'$^{s}$ unlike the case for the derivatives of $g =
\det{(g_{\mu\nu})}$.  This is well understood if one remembers that the
${{\hat{g}}_{4\mu}}$'$^{s}$, in the gauge theories point of view, are both
potentials and connections.}  ${\hat{g}}^{4\mu} = 0$, ${\hat{g}}^{44} =
1/{\hat{g}}_{44}$, $\hat{g} = \det{({\hat{g}}_{AB})} = g \,\,{\hat{g}}_{44}$,
${\hat{\Gamma}}^{\nu}_{\alpha\beta} = {\Gamma}^{\nu}_{\alpha\beta}$,
${\Gamma}^{\beta}_{\alpha\beta} = 0$, $\sqrt{ - \hat{g}} \, \,{\partial}_{\beta}
\,( \,{\hat{g}}^{4\beta} \,{\hat{\Gamma}}^{\nu}_{4\nu} \, ) = {\partial}_{\beta}
\,( \,\sqrt{ - \hat{g}} \, \,{\hat{g}}^{4\beta} \,{\hat{\Gamma}}^{\nu}_{4\nu} \,
)$, $\sqrt{ - \hat{g}} \, \,{\partial}_{\nu} \,( \,{\hat{g}}^{4\alpha}
\,{\hat{\Gamma}}^{\nu}_{4\alpha} \, ) = {\partial}_{\nu} \,( \,\sqrt{ - \hat{g}}
\, \,{\hat{g}}^{4\alpha} \,{\hat{\Gamma}}^{\nu}_{4\alpha} \, )$ and $(
\,{\partial}_{\nu} \sqrt{ - \hat{g}} \,) \,g^{\alpha\beta}
\,{\Gamma}^{\nu}_{\alpha\beta} = 0$ (recall that for any scalar quantity,
$\Omega$, one has identically ${\partial}_{\nu} \Omega \,( \,{\partial}_{\alpha}
g^{\nu\alpha} \,) = 0$).  In addition, the cylinder condition implies
${\hat{\Gamma}}^{4}_{44} = 0$.  So, one is left with the following expressions
\begin{equation} \label{G quantity 5D bis} \hat{G} = G + 2 g^{\alpha\beta}
\,{\hat{\Gamma}}^{\mu}_{\alpha4} \,{\hat{\Gamma}}^{4}_{\beta\mu} +
g^{\alpha\beta} \,{\hat{\Gamma}}^{4}_{\alpha4} \,{\hat{\Gamma}}^{4}_{\beta4} +
{\hat{g}}^{44} \,{\hat{\Gamma}}^{\mu}_{4\nu} \,{\hat{\Gamma}}^{\nu}_{4\mu} +
{\hat{g}}^{44} \,{\hat{\Gamma}}^{\mu}_{44} \,{\hat{\Gamma}}^{4}_{4\mu}
\end{equation} and, dropping the total divergence terms ${\partial}_{\beta} \,(
\,\sqrt{ - \hat{g}} \, \,{\hat{g}}^{4\beta} \,{\hat{\Gamma}}^{\nu}_{4\nu} \, )$
and ${\partial}_{\nu} \,( \,\sqrt{ - \hat{g}} \, \,{\hat{g}}^{4\alpha}
\,{\hat{\Gamma}}^{\nu}_{4\alpha} \, )$, \begin{equation} \label{pseudodivergence
bis} \hat{D} = \sqrt{{\hat{g}}_{44}} \,D \,- \,[ \,\frac{({\partial}_{\nu}
\sqrt{{\hat{g}}_{44}})}{\sqrt{\hat{g}_{44}}} \,{\hat{g}}^{44}
\,{\hat{\Gamma}}^{\nu}_{44} \, - \,{\partial}_{\nu} ({\hat{g}}^{44}
\,{\hat{\Gamma}}^{\nu}_{44}) \,+ \,\frac{({\partial}^{\alpha}
\sqrt{{\hat{g}}_{44}})}{\sqrt{{\hat{g}}_{44}}} \,{\hat{\Gamma}}^{4}_{\alpha4}
\,+ \,{\partial}_{\beta}(g^{\alpha\beta} {\hat{\Gamma}}^{4}_{\alpha4}) \,]
\,\sqrt{ - \hat{g}}, \end{equation} where the quantities \begin{equation}
\label{G quantity 4D} G = g^{\alpha\beta} \,( \,{\Gamma}^{\mu}_{\alpha\nu}
\,{\Gamma}^{\nu}_{\beta\mu} - {\Gamma}^{\nu}_{\alpha\beta}
\,{\Gamma}^{\mu}_{\nu\mu} \,) \end{equation} and \begin{equation}
\label{divergence 4D} D = {\partial}_{\nu} (\sqrt{ - g} \,{g}^{\alpha\beta}
\,{\Gamma}^{\nu}_{\alpha\beta}) - {\partial}_{\beta} (\sqrt{ - g}
\,{g}^{\alpha\beta} \,{\Gamma}^{\nu}_{\alpha\nu}).  \end{equation} are
respectively the spacetime analogous of $\hat{G}$ and $\hat{D}$.  Similarly, the
spacetime analogous of relation (\ref{EinsteinHilbert 5D}) reads
\begin{equation} \label{EinsteinHilbert 4D} \sqrt{- g} \,R = \sqrt{- g} \,G + D.
\end{equation} As one can see whereas $D$ is a total divergence of spacetime,
this is not the case for $\hat{D}$ since the terms involving the partial
derivatives with respect to the fith coordinate are missing because of the
cylinder condition.  At this point, the only Christoffel symbols we have to
compute explicitly are these of the form ${\hat{\Gamma}}^{4}_{\alpha4}$,
${\hat{\Gamma}}^{\mu}_{44}$, ${\hat{\Gamma}}^{\mu}_{4\nu}$ and
${\hat{\Gamma}}^{4}_{\beta\mu}$.  One gets the well known relations
\begin{equation} \label{Gamma1} {\hat{\Gamma}}^{4}_{\alpha4} = \frac{1}{2}
\,{\hat{g}}^{44} \,{\partial}_{\alpha} \,{\hat{g}}_{44}, \end{equation}
\begin{equation} \label{Gamma2} {\hat{\Gamma}}^{\alpha}_{44} = - \frac{1}{2}
\,{\partial}^{\alpha} \,{\hat{g}}_{44}, \end{equation} \begin{equation}
\label{Gamma3} {\hat{\Gamma}}^{\mu}_{4\nu} = \frac{1}{2} \,g^{\mu\alpha} \,(
\,{\partial}_{\nu} \,{\hat{g}}_{4\alpha} - {\partial}_{\alpha}
\,{\hat{g}}_{4\nu} \,) \end{equation} and \begin{equation} \label{Gamma4}
{\hat{\Gamma}}^{4}_{\beta\mu} = \frac{1}{2} \,{\hat{g}}^{44} \,(
\,{\partial}_{\mu} \,{\hat{g}}_{\beta4} + {\partial}_{\beta} \,{\hat{g}}_{4\mu}
\,).  \end{equation} Hence, it comes \begin{equation} \label{corrolaire 1}
{\hat{\Gamma}}^{\nu}_{4\nu} = 0 \end{equation} and \begin{equation}
\label{corrolaire 2} {\partial}_{\nu} ({\hat{g}}^{44}
\,{\hat{\Gamma}}^{\nu}_{44}) = - \,\frac{1}{2} \,( \,{\partial}_{\nu}
{\hat{g}}^{44} \,{\partial}^{\nu} {\hat{g}}_{44} \,+ \,{\hat{g}}^{44}
\,{\partial}_{\nu} \,{\partial}^{\nu} {\hat{g}}_{44} \,).  \end{equation}
Moreover, one checks easily that \begin{equation} \label{cancelation1}
g^{\alpha\beta}\,{\hat{\Gamma}}^{\mu}_{\alpha4} \,{\hat{\Gamma}}^{4}_{\beta\mu}
= 0 \end{equation} and \begin{equation} \label{cancelation2} g^{\alpha\beta} \,
{\hat{\Gamma}}^{4}_{\alpha4} \, {\hat{\Gamma}}^{4}_{\beta4} + {\hat{g}}^{44} \,
{\hat{\Gamma}}^{\mu}_{44} \, {\hat{\Gamma}}^{4}_{4\mu} = 0.  \end{equation} Thus
relations (\ref{G quantity 5D bis}) and (\ref{pseudodivergence bis}) reduce
respectively to \begin{equation} \label{G quantity 5D ter} \hat{G} = G +
\frac{1}{4} \,{\hat{g}}^{44} \,g^{\mu\alpha} \,g^{\nu\beta} \,(
\,{\partial}_{\nu} \,{\hat{g}}_{4\alpha} - {\partial}_{\alpha}
\,{\hat{g}}_{4\nu} \,) \,( \,{\partial}_{\mu} \,{\hat{g}}_{4\beta} -
{\partial}_{\beta} \,{\hat{g}}_{4\mu} \,) \end{equation} and \begin{equation}
\label{pseudodivergence ter} \hat{D} = \sqrt{{\hat{g}}_{44}} \,D - [
\,{\partial}_{\nu} \,({\hat{g}}^{44} \,{\partial}^{\nu} {\hat{g}}_{44} \,) \,]
\,\sqrt{ - \hat{g}}.  \end{equation} Hence, replacing relations (\ref{G quantity
5D ter}) and (\ref{pseudodivergence ter}) in relation (\ref{EinsteinHilbert 5D})
yields on account of relation (\ref{EinsteinHilbert 4D}) :  \begin{equation}
\label{Ricci 5D bis} \hat{R} = R + \frac{1}{4} \,{\hat{g}}^{44} \,g^{\mu\alpha}
\,g^{\nu\beta} \,( \,{\partial}_{\nu} \,{\hat{g}}_{4\alpha} -
{\partial}_{\alpha} \,{\hat{g}}_{4\nu} \,) \,( \,{\partial}_{\mu}
\,{\hat{g}}_{4\beta} - {\partial}_{\beta} \,{\hat{g}}_{4\mu} \,) -
\,{\partial}_{\nu} \,({\hat{g}}^{44} \,{\partial}^{\nu} {\hat{g}}_{44}).
\end{equation} Now, the above relation is equivalent to the following
\begin{equation} \label{Ricci 5D ter} \hat{R} = R + \frac{1}{4\,{\hat{g}}_{44}}
\,g^{\mu\alpha} \,g^{\nu\beta} \,( \,{\partial}_{\nu} \,{\hat{g}}_{4\alpha} -
{\partial}_{\alpha} \,{\hat{g}}_{4\nu} \,) \,( \,{\partial}_{\mu}
\,{\hat{g}}_{4\beta} - {\partial}_{\beta} \,{\hat{g}}_{4\mu} \,) + \frac{1}{2}
\,\frac{{\partial}_{\nu} {\hat{g}}_{44} \,{\partial}^{\nu}
{\hat{g}}_{44}}{({\hat{g}}_{44})^{2}} \end{equation} since \begin{eqnarray*}
\sqrt{ - \hat{g}} \,{\partial}_{\nu} \,({\hat{g}}^{44} \,{\partial}^{\nu}
{\hat{g}}_{44}) & = & {\partial}_{\nu} \,(\sqrt{ - \hat{g}} \,{\hat{g}}^{44}
\,{\partial}^{\nu} {\hat{g}}_{44}) - {\hat{g}}^{44} \,({\partial}^{\nu}
{\hat{g}}_{44}) \,({\partial}_{\nu} \,\sqrt{ - \hat{g}}) \\ \\ & & \mbox{} =
{\partial}_{\nu} \,(\sqrt{ - \hat{g}} \,{\hat{g}}^{44} \,{\partial}^{\nu}
{\hat{g}}_{44}) - \frac{1}{2} \,\frac{{\partial}_{\nu} {\hat{g}}_{44}
\,{\partial}^{\nu} {\hat{g}}_{44}}{({\hat{g}}_{44})^{2}} \,\sqrt{ - \hat{g}}
\end{eqnarray*} and we may drop the total divergence for our concern.  Clearly,
the sign accompanying the kinetic term as regards the fifteen degree of freedom
(identified to the scalar field, up to a power-law), ${\hat{g}}_{44}$, of the
Kaluza-Klein theory is unambigously negative.\\\\In Jordan-Fierz frame, the
fields potentials are defined by :  ${\hat{g}}_{44} = {\phi}^{2}$,
${\hat{g}}_{4\mu} = \kappa \,{\phi}^{2} \,A_{\mu}$ and ${\hat{g}}_{\mu\nu} =
g_{\mu\nu} + {\kappa}^{2} \,{\phi}^{2} \,A_{\mu} \,A_{\nu}$.  Carrying these
relations into relation (\ref{Ricci 5D ter}) yields expression (\ref{KK action
Jordan}) of the five dimensional Einstein-Hilbert action.  In Einstein-Pauli
frame, the potentials of the fields are defined by :  ${\hat{g}}_{44} =
{\tilde{\phi}}^{2/3}$, ${\hat{g}}_{4\mu} = \kappa \,{\tilde{\phi}}^{2/3}
\,A_{\mu}$ and ${\hat{g}}_{\mu\nu} = {\tilde{\phi}}^{-1/3} \,(
\,{\tilde{g}}_{\mu\nu} + {\kappa}^{2} \,\tilde{\phi} \,A_{\mu} \,A_{\nu})$.
Carrying these relations into relation (\ref{Ricci 5D ter}) yields expression
(\ref{KK action Pauli}) of the five dimensional Einstein-Hilbert action.

\end{document}